\documentclass[epj]{svjour}
\usepackage{amsmath,amssymb,amsfonts,graphicx,cite,algorithmic,algorithm,multirow}
\graphicspath{{graphics/}}
\makeatletter
\ifx\input@path\@undefined
\def\input@path{{graphics/}}
\else
\g@addto@macro\input@path{{graphics/}}
\fi
\makeatother

\preprint{DESY 12-215\\ MCnet-12-13}

\title{Controlling inclusive cross sections in\\ parton shower + matrix element merging}

\author{Simon Pl\"atzer}

\institute{DESY, Notkestrasse 85, D-22607 Hamburg, Germany}

\date{\today}

\abstract{We propose an extension of matrix element plus parton shower
  merging at tree level to preserve inclusive cross sections obtained
  from the merged and showered sample. Implementing this constraint
  generates approximate next-to-leading order (NLO) contributions
  similar to the LoopSim approach. We then show how full NLO, or in
  principle even higher order, corrections can be added consistently,
  including constraints on inclusive cross sections to account for yet
  missing parton shower accuracy at higher logarithmic order. We also
  show how NLO accuracy below the merging scale can be obtained.
   }

\begin{document}

\onecolumn

\maketitle


\section{Introduction}

Parton shower event generators,
\cite{Sjostrand:2007gs,Bahr:2008pv,Gleisberg:2003xi}, to name only few
of the most recent general-purpose simulations, are by now
indispensable workhorses of experimental as well as theoretical
studies for comparing (standard model) theoretical predictions to
measured observables in a most detailed way. Within recent years,
tremendous progress has been made in improving the approximations
underlying those simulations by exact calculations in perturbation
theory. Starting from simple matrix element corrections on the one
hand, improving the hardest parton shower emission to be driven by the
exact tree level matrix element \cite{Seymour:1994df,Norrbin:2000uu},
subsequently tree level matrix elements for $n>1$ hardest emissions
have been included consistently
\cite{Catani:2001cc,Lonnblad:2001iq,Krauss:2002up,Hoche:2006ph,
  Lavesson:2007uu,Hoeche:2009rj,Hamilton:2009ne,Lonnblad:2011xx}. These
approaches of {\it merging} multileg matrix elements at tree level,
{\it i.e.}  leading order (LO), have been accompanied by efforts of
{\it matching} parton shower simulations and perturbative calculations
at next-to-leading order (NLO), including both matrix elements for
additional hard emission at tree level as well as virtual, one-loop
corrections to the hard process of interest,
\cite{Dobbs:2001dq,Frixione:2002ik,
  Nason:2004rx,Nagy:2005aa,Frixione:2010ra,Platzer:2011bc}. On the
other hand, parton shower algorithms themselves have evolved from
crude approximations of multiple parton emission to more and more
refined and precise tools
\cite{Nagy:2006kb,Winter:2007ye,Dinsdale:2007mf,Schumann:2007mg,Giele:2007di,Platzer:2009jq,Kilian:2011ka},
eventually aiding attempts of combining fixed-order calculations with
shower resummation in a most consistent way.

NLO matching has been used to improve tree level merging algorithms,
\cite{Bauer:2008qh,Bauer:2008qj,Hamilton:2010wh,Hoche:2010kg}, but
only recently, taking full advantage of developments in both merging,
matching and shower algorithms, first experience has been gained in
combining NLO QCD corrections to multijet final states of different
multiplicity with subsequent parton shower emissions
\cite{Lavesson:2008ah,Hoeche:2012yf,Gehrmann:2012yg,Frederix:2012ps}. The
aim of such approaches is to obtain an event sample which will provide
LO (NLO) precision for observables controlled by hard $n$-parton
emission as long as exact QCD corrections are included for up to $n$
partons at tree level ($n$ partons at one loop and $n+1$ partons at
tree level), while including resummation by means of shower simulation
at whatever accuracy is provided by these algorithms.

While merging approaches at tree level have been proven to be
amazingly powerful in describing experimental data, and can now be
considered to be well understood at a theoretical level (particularly
with respect to logarithmic dependence on the merging scale separating
hard, matrix element driven emissions from softer parton shower
emissions), their generalization to include NLO corrections yet
suffers from impact of the merging scale at a level of logarithmic
accuracy which by no means can be provided by existing shower
algorithms. The most striking signs of this dependence are expected in
inclusive cross sections as predicted by such simulations, and
proposals to cure the problem by explicit input of resummation results
at higher accuracy have been made \cite{Tackmann:Geneva}. In the present
contribution we try to approach this problem from a rather pragmatic
point of view by setting up a formalism to systematically include
higher oder corrections in parton shower simulations, while satisfying
constraints to obtain the proper inclusive cross sections at the
respective order of perturbative calculations available to the merging
algorithm.

\section{Setting the scene}
\label{secs:settingthescene}

We will consider a generic parton shower with an evolution variable
$q$ (all other kinematic variables of a splitting are suppressed for
the sake of readability), which is driven by splitting kernels
$P(\phi_n,q)$. Here and in the following, $\phi_n$ denotes a partonic
configuration (phase space point) containing $n$ additional partons
with respect to the lowest order process of interest, and
$P(\phi_n,q)$ determines the dynamics of emission at a scale $q$ off
the partonic system $\phi_n$. The Sudakov form factor associated to
$P(\phi_n,q)$, evolving from a hard scale $Q$ to a soft scale $q$, is
given by
\begin{equation}
\Delta_n(q|Q) = \exp\left(-\int_q^Q {\rm d}k\ \frac{{\rm d}\phi_{n+1}}{{\rm d}\phi_{n}{\rm d}k}\ P(\phi_n,k)\right) \ ,
\end{equation}
with the phase space Jacobian ${\rm d}\phi_{n+1}/{\rm d}\phi_{n}$
depending implicitly on the shower kinematic variables.  Partonic
configurations $\phi_n$ are determined according to differential cross
sections ${\rm d}\sigma(\phi_n,q_n|\cdots|q_0)$, where the sequence of
scales $q_n < q_{n-1}, ... , q_0$ is either directly determined from
shower evolution, or is assigned by a clustering procedure
corresponding to the inverse of a possible shower evolution $\phi_0\to
\phi_n$, if the cross section is determined by exact matrix
elements.\footnote{Note that ordered histories may not always
  exist. In these cases we can assume that, for example, the history
  with the longest ordered history from smaller to larger scales has
  been chosen. This will not pose a problem for the formalism outlined
  here, provided we set Sudakov form factors with unordered scales
  equal to one, $\Delta_n(Q|q)=1$ for $Q>q$. Also note that a
  clustering down to the lowest order process may not always be
  possible, in case of which we terminate the clustering sequence at
  the last possible clustering step.} In the case of exact matrix
elements, reweighting to account for couplings and PDF factors
evaluated at the clustering scales may as well be included in this
notation. For the sake of readability we will skip the whole sequence
of scales and denote ${\rm d}\sigma(\phi_n,q_n|...|q_0) = {\rm
  d}\sigma(\phi_n,q_n)$. The parton shower action on cross sections
for $\phi_n$ events is given by\footnote{For the following note that
  ${\rm PS}[\cdot]$ is linear.}
\begin{equation}
{\rm PS}_\mu\left[ {\rm d}\sigma(\phi_n,q_n)\right] =
{\rm d}\sigma(\phi_n,q_n) \Delta_n(\mu|q_n) +
{\rm PS}_\mu\left[{\rm d}\sigma(\phi_n,q_n)
\frac{{\rm d}\phi_{n+1}}{{\rm d}\phi_{n}}P_\mu(\phi_n,q_{n+1})\Delta_n(q_{n+1}|q_n)\right]\ .
\end{equation}
The first contribution are events where no further radiation has been
generated ($\mu$ denotes the parton shower cutoff and the splitting
kernels $P_\mu(\phi_{n},q_{n+1}) =
P(\phi_{n},q_{n+1})\theta(q_{n+1}-\mu)$ vanish for scales below the
cutoff), while the second contribution corresponds to at least one
emission.

Considering parton showering off the lowest order tree level cross
section, ${\rm d}\sigma^{(0)}(\phi_0,q_0)$, we find that the first $N$
iterations of the parton shower action give rise to
\begin{multline}
\label{eqs:pscrosssections}
{\rm PS}_\mu\left[{\rm d}\sigma^{(0)}(\phi_0,q_0) \right] =
{\rm d}\sigma^{(0)}(\phi_0,q_0)
\sum_{k=0}^{N-1} 
\frac{{\rm d}\phi_{k}}{{\rm d}\phi_{0}}P_\mu(\phi_{k-1},q_{k})\cdots
P_\mu(\phi_0,q_{1}) \Delta_{k}(\mu|q_k|\cdots|q_0)\ + \\
{\rm PS}_\mu\left[   
{\rm d}\sigma^{(0)}(\phi_0,q_0)\frac{{\rm d}\phi_{N}}{{\rm d}\phi_{0}}P_\mu(\phi_{N-1},q_{N})\cdots
P_\mu(\phi_0,q_{1}) \Delta_{N-1}(q_{N}|\cdots|q_0)\right] \ ,
\end{multline}
where $\Delta_{k}(\mu|q_k|\cdots|q_0) =
\Delta_{k}(\mu|q_k)\cdots\Delta_0(q_1|q_0)$. Considering the exclusive
and inclusive cross sections for producing exactly $n$ or at least $n$
partons, respectively, we have
\begin{eqnarray}\nonumber
= n &\qquad& {\rm d}\sigma^{(0)}(\phi_0,q_0)
\frac{{\rm d}\phi_{n}}{{\rm d}\phi_{0}}P_\mu(\phi_{n-1},q_{n})\cdots
P_\mu(\phi_0,q_{1}) \Delta_{n}(\mu|q_{n}|\cdots|q_0)\\\nonumber
\ge n & \qquad & {\rm d}\sigma^{(0)}(\phi_0,q_0)
\frac{{\rm d}\phi_{n}}{{\rm d}\phi_{0}}P_\mu(\phi_{n-1},q_{n})\cdots
P_\mu(\phi_0,q_{1}) \Delta_{n-1}(q_{n}|\cdots|q_0) \ .
\end{eqnarray}
Note that the latter expression is a direct implication of the parton
shower being unitary. In particular, the total inclusive cross section
is not altered by the parton shower and determined by the lowest order
`input' cross section ${\rm d}\sigma^{(0)}$, as would be expected from
the very definition of an inclusive cross section.

\begin{figure}
\begin{center}
\includegraphics[scale=0.5]{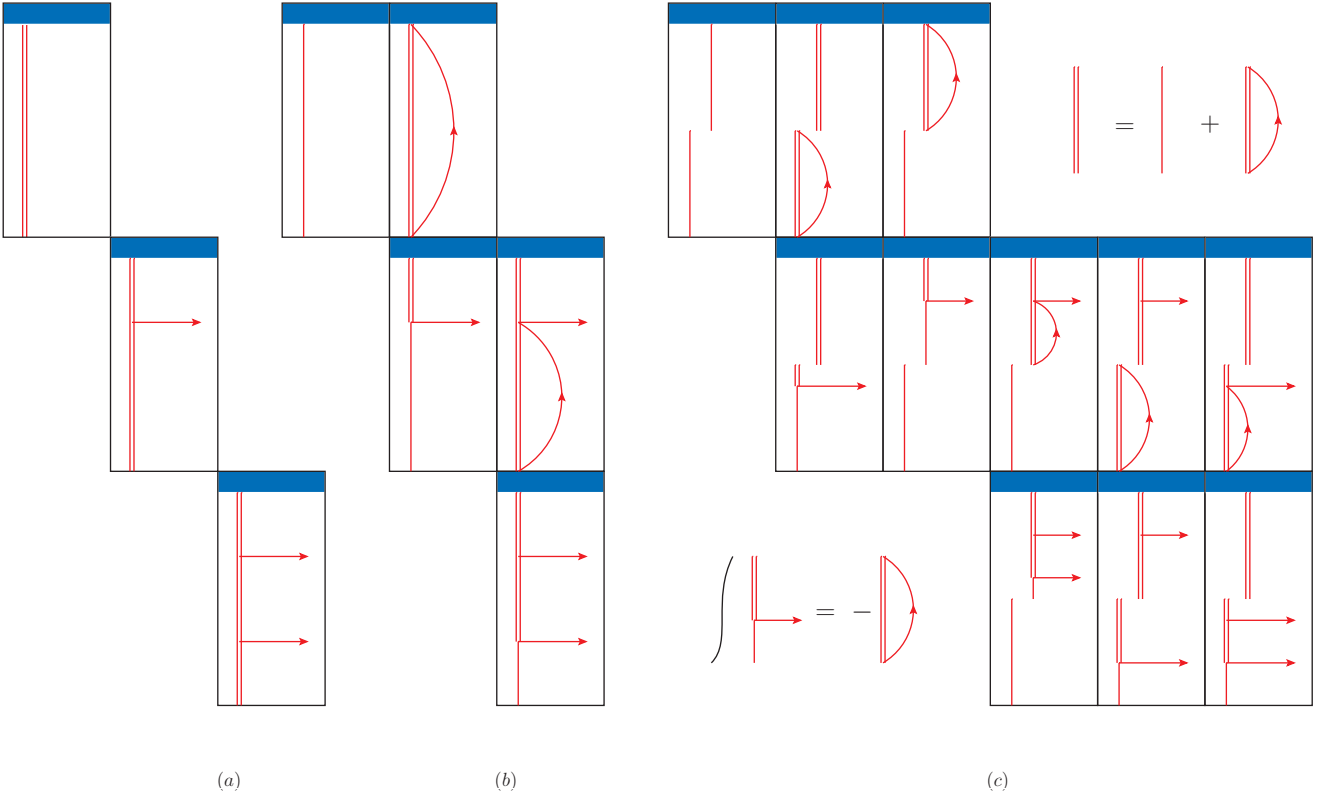}
\end{center}
\caption{\label{figs:showerdiags}Main properties of parton showers
  illustrated in a diagrammatic way. Each box with a black frame
  denotes a $n$-parton contribution, where $n$ runs from top to
  bottom. Rows need to be summed to obtain the total cross section
  driving a given parton multiplicity. The parton shower evolves along
  the vertical red lines from larger (top) to smaller (bottom)
  scales. Double red lines denote the Sudakov factors associated to a
  given evolution interval, horizontal arrows branching off denote
  emission at a certain scale. With the parton shower cutoff at the
  bottom of each box, part (a) shows exclusive cross sections for
  zero, one and two partons emitted,
  cf. eq.~\ref{eqs:pscrosssections}. Part (b) illustrates the shower
  cross section integrated over the contributions of more than two
  emissions, making use of eq.~\ref{eqs:unitarity}; the form of
  inclusive cross sections is obvious, as each column of two boxes
  integrates to zero. Part (c) illustrates how the shower splits up
  into two evolution intervals, cf. eq.~\ref{eqs:scaleinterval}.}
\end{figure}

This property is a direct result of the fact that the emission
contribution of the parton shower at scale $q$ is an exact
differential of the no emission probability down to $q$,
\begin{equation}
\label{eqs:unitarity}
\int_q^{q_{k-1}}{\rm d}q_k  \frac{{\rm d}\phi_{k}}{{\rm d}\phi_{k-1}{\rm d}q_k}P(\phi_{k-1},q_{k})\Delta_{k-1}(q_k|q_{k-1}) =
1 - \Delta_{k-1}(q|q_{k-1}) \ .
\end{equation}
Introducing $P_{\mu |
  \rho}(\phi_{k-1},q_k)=P(\phi_{k-1},q_k)\theta(\rho
-q_k)\theta(q_k-\mu)$ with $\rho > \mu$, and associating ${\rm
  PS}_{\mu|\rho}$ as the associated action, we also have
\begin{equation}
\label{eqs:scaleinterval}
{\rm PS}_\mu[\cdot ] =  {\rm
  PS}_{\mu|\rho}\left[{\rm PS}_\rho[\cdot ]\right] \ ,
\end{equation}
{\it i.e.} we can always split the shower evolution into two (or more)
scale ranges. We illustrate the main formulae discussed in this
section in a diagramatic manner in fig.~\ref{figs:showerdiags}.

\section{Tree-level merging}
\label{secs:mergingcondition}

Having analyzed the exclusive and inclusive cross sections for
$n$-parton production by the shower, we will now review how these
predictions can be improved by including higher multiplicity
tree-level matrix elements. Clearly, the desired result in terms of
exclusive $n$-parton states is to replace the product of splitting
kernels approximating an exact tree-level matrix element by the full
answer. To this extent, in the presence of a maximum number of $N$
additional partons to be described by matrix elements, we are
therefore to satisfy a {\it merging condition}
\begin{equation}
\label{eqs:mergeingcondition}
{\rm PS}_\mu\left[{\rm d}\sigma^{\text{merged}}_{N,\mu}\right] =
\sum_{k=0}^{N-1} {\rm d}\sigma_\mu^{(0)}(\phi_k,q_k)\Delta_{k}(\mu|q_k|\cdots |q_0)\ +
{\rm PS}_\mu\left[ {\rm d}\sigma_\mu^{(0)}(\phi_N,q_N)\Delta_{N-1}(q_{N}|\cdots |q_0) \right] \ .
\end{equation}
As is the case for plain parton shower emissions, there are no
emissions with a scale below the shower cutoff $\mu$, which we here
apply to regulate the divergences in the additional tree-level matrix
elements considered. For this reason we have introduced ${\rm
  d}\sigma_\mu^{(0)}(\phi_n,q_n)={\rm
  d}\sigma^{(0)}(\phi_n,q_n)\theta(q_n-\mu)$.

It is straightforward to show that the merging condition can be satisfied by
\begin{equation}
{\rm d}\sigma^{\text{merged}}_{N,\mu} =
{\rm d}\sigma(\phi_0,q_0) +
\sum_{k=1}^{N} \left({\rm d}\sigma_\mu^{(0)}(\phi_k,q_k) - 
\frac{{\rm d}\phi_k}{{\rm d}\phi_{k-1}}P_\mu(\phi_{k-1},q_k){\rm d}\sigma_\mu^{(0)}(\phi_{k-1},q_{k-1})\right) 
\Delta_{k-1}(q_k|\cdots|q_0) \ .
\end{equation}
Note that we are free to introduce a merging scale $\rho > \mu$ by acting the
parton shower on ${\rm d}\sigma^{\text{merged}}_{N,\rho}$. The result is similar to
eq.~\ref{eqs:mergeingcondition}, upon replacing
\begin{equation}
{\rm d}\sigma_\mu^{(0)}(\phi_k,q_k)\quad\to\quad 
{\rm d}\sigma_\rho^{(0)}(\phi_k,q_k) +
\sum_{l=1}^k P_{\mu | \rho}(\phi_{k-1},q_{k})\cdots P_{\mu | \rho}(\phi_{k-l},q_{k-l+1})
\frac{{\rm d}\phi_k}{{\rm d}\phi_{k-l}}{\rm d}\sigma_\rho^{(0)}(\phi_{k-l},q_{k-l}) \ ,
\end{equation}
{\it i.e.} emissions at scales smaller than $\rho$ are purely
driven by shower dynamics, whereas above the exact tree level matrix
element is used. The merging scale at this
point can be viewed as just an efficiency tweak for not having to
evaluate exact matrix elements in a region where they are well
approximated by the shower.

For later purposes, and to make the connection to CKKW-type merging
\cite{Catani:2001cc,Lonnblad:2001iq,Krauss:2002up,Hoche:2006ph,
  Lavesson:2007uu,Hoeche:2009rj,Hamilton:2009ne,Lonnblad:2011xx}
explicit, we rewrite the shower cross section in presence of the
merging scale $\rho$ as
\begin{equation}
{\rm PS}_\mu\left[{\rm d}\sigma^{\text{merged}}_{N,\rho}\right] =
\sum_{k=0}^{N-1} {\rm PS}_{\mu | \rho}\left[{\rm d}\sigma_\rho^{(0)}(\phi_k,q_k)\Delta_{k}(\rho|q_k|\cdots |q_0)\right]
+{\rm PS}_\mu\left[ {\rm d}\sigma_\rho^{(0)}(\phi_N,q_N)\Delta_{N-1}(q_{N}|\cdots |q_0) \right] \ .
\end{equation}
Note that, except
for the highest matrix element multiplicity $N$ present, shower
emissions are confined to happen below the merging scale.
Using eq.~\ref{eqs:scaleinterval}, the previous result is simply given by
\begin{equation}
\label{eqs:mergedmerged}
{\rm PS}_\mu\left[{\rm d}\sigma^{\text{merged}}_{N,\rho}\right] =
{\rm PS}_{\mu | \rho}\left[
\sum_{k=0}^{N-1} {\rm d}\sigma_\rho^{(0)}(\phi_k,q_k)\Delta_{k}(\rho|q_k|\cdots |q_0)
+{\rm PS}_\rho\left[ {\rm d}\sigma_\rho^{(0)}(\phi_N,q_N)\Delta_{N-1}(q_{N}|\cdots |q_0) \right]\right] \ .
\end{equation}
In the merged sample, cross sections for exclusive $n\le N$ parton
configurations above the merging scale $\rho$ are determined by the
respective tree level matrix elements including the proper Sudakov
suppression, while exclusive cross sections for $n$ partons down to the
shower cutoff are determined by a mixture of matrix elements and
shower splitting functions depending on which part of the relevant
scale sequence is below or above the merging scale. The proper Sudakov
suppression is as well retained in the latter case.

Note that the merged cross section and its showered counterpart,
eq.~\ref{eqs:mergedmerged}, cover several approaches to tree level
merging so long as the merging resolution coincides with the parton
shower resolution in case of which no issues with truncated/vetoed
showers \cite{Hamilton:2009ne,Hoeche:2009rj} do appear. Throughout the
paper we will assume that this is always the case; a mismatch between
merging and shower resolution does not pose a conceptual problem to
the algorithm proposed in the following. In the presence of a merging
scale $\rho$, the formalism outlined here covers the standard merging
algorithms of reweighted tree level matrix elements with showering
below the merging scale. In this case, upon putting $P_\rho\to 0$
except for emissions off the highest multiplicity, the shower
subtractions are absent in ${\rm
  d}\sigma^{\text{merged}_{N,\rho}}$. If no merging scale is present,
iterated matrix element corrections are contained in the master
formula upon replacing the shower splitting kernels for emissions off
a system of up to $N-1$ partons by the respective ratio of tree level
matrix elements. In this case, the merged cross section coincides with
the lowest order cross section. To work in a most generic setup, we
will stay with the general solution to the merging condition, which
does not require any modification to the shower acting downstream the
hard processes obtained from the merged cross section.

For convenience, let us introduce the functional
\begin{equation}
\label{eqs:inverseps}
{\rm PS}^{-1}_\mu\left[{\rm d}\sigma_\mu(\phi_n,q_n)\right] =
\frac{{\rm d}\sigma_\mu(\phi_n,q_n)}{\Delta_n(\mu|q_n)} -
\frac{{\rm d}\phi_{n+1}}{{\rm d}\phi_{n}} P_\mu(\phi_n,q_{n+1})
\frac{{\rm d}\sigma_\mu(\phi_n,q_n)}{\Delta_n(\mu|q_{n+1})}
\end{equation}
which satisfies
\begin{equation}
{\rm PS}_\mu\left[ {\rm
  PS}^{-1}_\mu[{\rm d}\sigma_\mu(\phi_n,q_n)] \right] = {\rm d}\sigma_\mu(\phi_n,q_n)
\qquad\text{and}\qquad 
{\rm PS}_\mu\left[ {\rm
  PS}^{-1}_\rho[{\rm d}\sigma_\rho(\phi_n,q_n)] \right] =
{\rm PS}_{\mu | \rho}\left[{\rm d}\sigma_\rho(\phi_n,q_n)\right] \ .
\end{equation}
Using this definition, we can rewrite the merged cross section as
\begin{equation}
{\rm d}\sigma^{\text{merged}}_{N,\rho} =
\sum_{k=0}^{N-1}{\rm PS}_\rho^{-1}\left[ {\rm d}\sigma_\rho^{(0)}(\phi_k,q_k) \Delta_k(\rho|q_k|\cdots|q_0) \right]
+{\rm d}\sigma_\rho^{(0)}(\phi_N,q_N)\Delta_{N-1}(q_N|\cdots |q_0) \ ,
\end{equation}
which summarizes the solution to the tree level merging condition in a
most transparent way. Being of purely formal use for the
present letter, it is worth noting that ${\rm PS}_\rho^{-1}$
facilitates the subtraction of parton shower contributions above the
scale $\rho$ {\it at all orders}. Turning this observation around, it
can actually be used as a generating functional for subtractions
needed to satisfy fixed-order matching conditions upon expanding to
the respective order. Indeed,
\begin{multline}
{\rm PS}_\mu^{-1}\left[{\rm d}\sigma_\mu(\phi_n,q_n)\right] =\\
{\rm d}\sigma_\mu(\phi_n,q_n)\left(1 + \int_\mu^{q_n}{\rm d}q_{n+1} 
\frac{{\rm d}\phi_{n+1}}{{\rm d}\phi_{n}{\rm d}q_{n+1}} P_\mu(\phi_n,q_{n+1})\right)
-
\frac{{\rm d}\phi_{n+1}}{{\rm d}\phi_{n}} P_\mu(\phi_n,q_{n+1})
{\rm d}\sigma_\mu(\phi_n,q_n) + {\cal O}(\alpha_s^2)
\end{multline}
is readily identified as the corrections required for NLO
matching. Note also that the action of ${\rm PS}_\rho^{-1}$ can
actually be implemented in a Monte Carlo simulation by performing the
veto algorithm from smaller to larger scales while using a simple
prescription to exponentiate $-P(\phi_n,q)$ at the expense of
introducing weighted events \cite{dissertationsimon}.

\section{Inclusive cross sections}
\label{secs:inclusive}

Let us now turn to inclusive cross sections. We are free to split up
the phase space integration into a region of scales below the merging
scale $\rho$, and one above. Performing the integration over smaller
scales first (or, equivalently, considering jet cross sections in a jet
measure exactly corresponding to inverting the parton shower), we
effectively remove the shower action ${\rm PS}_{\mu|\rho}$ from
eq.~\ref{eqs:mergedmerged}.

Looking at inclusive cross sections of at least $n\le N$ jets,
we face essentially a non-unitary evolution due to the matrix elements
not coinciding with products of splitting kernels anymore. While for
the highest multiplicity we find the same pattern as present in the
parton shower by the very definition of the merging condition,
\begin{eqnarray}\nonumber
\ge N &\qquad& {\rm d}\sigma_\rho^{(0)}(\phi_N,q_N)\Delta_{N-1}(q_N|\cdots |q_0) \ ,
\end{eqnarray}
the result for one parton less does not anymore resemble the
functional dependence present in the shower:\footnote{We have
  added and subtracted the contribution from a shower emission off the
  $N-1$ parton state, where the positive term has been absorbed by the
  integration over the scale $q_N$ to yield the Sudakov form factor in
  the first line.}
\begin{eqnarray}\nonumber
\ge N-1 &\qquad& {\rm
  d}\sigma_\rho^{(0)}(\phi_{N-1},q_{N-1})\Delta_{N-2}(q_{N-1}|\cdots
|q_0) +\\\nonumber&&\int_\rho^{q_{N-1}} {\rm d}q_N \left( \frac{{\rm
    d}\sigma_\rho^{(0)}(\phi_N,q_N)}{{\rm d}q_N}-\frac{{\rm
    d}\phi_{N}}{{\rm d}\phi_{N-1}{\rm d}q_N}
P_\rho(\phi_{N-1},q_N){\rm d}\sigma_\rho^{(0)}(\phi_{N-1},q_{N-1})
\right)\Delta_{N-1}(q_N|\cdots|q_0) \ .
\end{eqnarray}

\begin{figure}
\begin{center}
\includegraphics[scale=0.5]{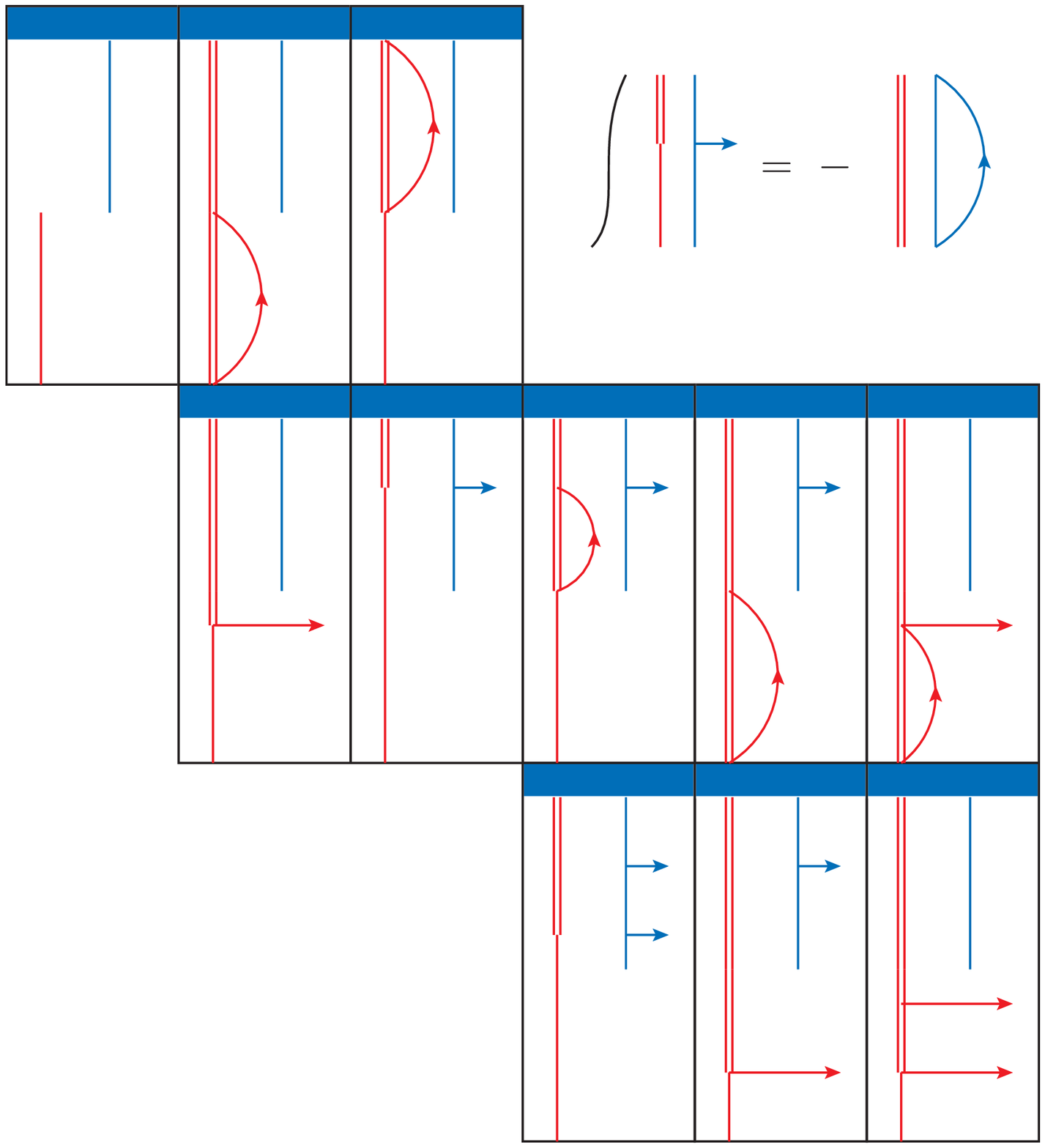}
\includegraphics[scale=0.5]{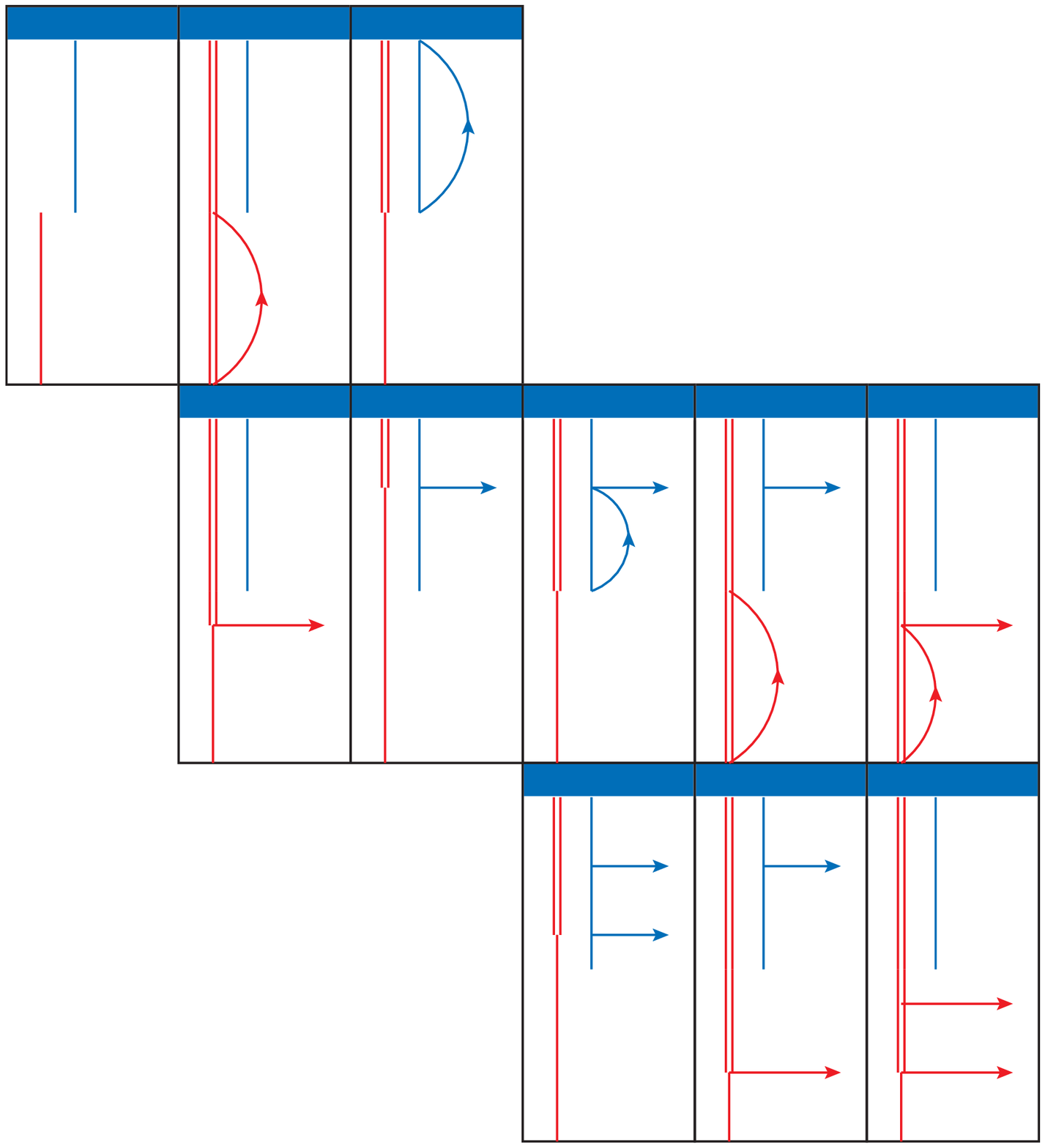}
\end{center}
\caption{\label{figs:merging} The left panel shows the solution to the
  merging condition for $N=2$, eq.~\ref{eqs:mergedmerged}, with
  inconsistencies in inclusive cross sections made explicit. The right
  panel shows the cross sections obtained from the merged sample upon
  including counter terms to restore inclusive cross sections as given
  by eq.~\ref{eqs:mergedinclusive}. Here, emissions calculated
  according to exact tree level matrix elements are made explicit by
  arrows branching off blue lines.}
\end{figure}

In order to restore the form of the cross section obtained by the
shower, we need to subtract the second term obtained for the inclusive
$N-1$ jet cross section. This pattern then continues, and one
possibility to solve the issues with inclusive cross sections is then
given by the replacement
\begin{multline}\label{eqs:mergedinclusive}
{\rm d}\sigma_{N,\rho}^{\text{merged}} \to \\
{\rm d}\sigma_{N,\rho}^{\text{merged}} - 
\sum_{k=0}^{N-1}
{\rm PS}_\rho^{-1}\left[
\int_\rho^{q_{k}} {\rm d}q_{k+1} \left( \frac{{\rm
  d}\sigma_\rho^{(0)}(\phi_{k+1},q_{k+1})}{{\rm d}q_{k+1}}-\frac{{\rm d}\phi_{k+1}}{{\rm
    d}\phi_{k}{\rm d}q_{k+1}} P(\phi_{k},q_{k+1}){\rm
  d}\sigma_\rho^{(0)}(\phi_{k},q_{k})
\right)\Delta_{k}(q_{k+1}|\cdots|q_0)
\right] \\
= \sum_{k=0}^{N-1} {\rm PS}_\rho^{-1}\left[ 
\left( {\rm d}\sigma_\rho^{(0)}(\phi_k,q_k)-\int_\rho^{q_k}{\rm d}q_{k+1}  
\frac{{\rm
  d}\sigma_\rho^{(0)}(\phi_{k+1},q_{k+1})}{{\rm d}q_{k+1}} \Delta_{k}(q_{k+1}|q_k)\right) \Delta_{k-1}(q_k|\cdots|q_0)
\right]\\
+{\rm d}\sigma_\rho^{(0)}(\phi_N,q_N)\Delta_{N-1}(q_N|\cdots |q_0) \ .
\end{multline}
The solution to the merging condition as discussed in
section~\ref{secs:mergingcondition}, as well as the results obtained
from the merged sample including corrections for inclusive cross
sections are depicted in fig.~\ref{figs:merging}.  One remark is in
order here: while we have now achieved to retain the expected form of
inclusive cross sections (including the constraint that the total
inclusive cross section is given by the lowest order input cross
section) the exclusive cross sections with $n\le N$ partons above the
merging scale take a seemingly different form,
\begin{equation}\nonumber
{\rm PS}_{\mu|\rho}\left[ \left( {\rm
    d}\sigma_\rho^{(0)}(\phi_n,q_n) - \int_\rho^{q_n}{\rm d}q_{n+1}
  \frac{{\rm d}\sigma_\rho^{(0)}(\phi_{n+1},q_{n+1})}{{\rm
      d}q_{n+1}}\Delta_n(q_{n+1}|q_n) \right) \Delta_{n-1}(q_n|\cdots|q_0)\right] \ .
\end{equation}
It is clear that such a change will occur when trying to restore
inclusive cross sections. We stress however that the form encountered
here does not pose a problem in terms of logarithmic accuracy, as long
as the shower kernels are truly capable of reproducing the
singly-unresolved limits of the tree level matrix elements. In this
case the dominant contributions of integral in the second term will
render those singularities to precisely integrate to ${\rm
  d}\sigma_\rho^{(0)}(\phi_n,q_n)(1-\Delta_n(\rho|q_n))$, restoring
the expected Sudakov suppression.

\section{Injecting NLO corrections}
\label{secs:nlo}

Having improved the parton shower by tree level matrix elements,
including constraints on inclusive cross sections, we can now turn to
the case of including NLO corrections, which we assume are here available
up to $M$ additional partons with respect to the lowest order process.

Before discussing this in greater detail, let us consider the
respective contributions obtained from the merged sample with
corrections for inclusive cross sections. The cross sections to be
analyzed are the contributions of exactly $n$ partons above the
merging scale (integrated over phase space below $\rho$), which we will
have to correct to be driven by the respective NLO cross sections.
The exclusive cross sections for $n$ partons above the merging scale
are given by
\begin{equation}
{\rm d}\sigma_{n,\text{excl}}^{\text{nLO}} =
 \Delta_{n-1}(q_n|\cdots|q_0) \left( {\rm
    d}\sigma_\rho^{(0)}(\phi_n,q_n) - \int_\rho^{q_n}{\rm d}q_{n+1}
  \frac{{\rm d}\sigma_\rho^{(0)}(\phi_{n+1},q_{n+1})}{{\rm
      d}q_{n+1}}\Delta_n(q_{n+1}|q_n) \right) \ .
\end{equation}
Note that this does contain the Sudakov form factor for obtaining at
least $n$ partons above the merging scale. By the discussion in the
previous section, the contribution in brackets will also provide for
the remaining Sudakov factor down to the merging scale. Note also,
that upon expanding the contribution in brackets to first order in
$\alpha_s$, we obtain the approximated, exclusive NLO cross section
for $n$ jets above a resolution $\rho$ as given by the
\textsf{LoopSim} prescription\cite{Rubin:2010xp}, which motivated the label nLO.

This observation directly implies that, if we add a correction cross
section given by the inclusive NLO corrections,
$$ {\rm PS}_\rho^{-1}\left[ \left({\rm d}\sigma_\rho^{(1)}(\phi_n,q_n)
  + \int_0^{q_n}{\rm d}q_{n+1}\frac{{\rm
      d}\sigma^{(0)}(\phi_{n+1},q_{n+1})}{{\rm
      d}q_{n+1}}\theta(q_n-\rho)\right)\Delta_{n-1}(q_n|\cdots|q_0)
  \right] \ ,
$$ we will obtain NLO accuracy for exclusive $n$ parton configurations
above the merging scale,
\begin{equation}
\label{eqs:mergematch}
{\rm d}\sigma_{n,\text{excl}}^{\text{NLO}} = \Delta_{n-1}(q_n|\cdots|q_0)
 \left( {\rm d}\sigma_\rho^{(0)}(\phi_n,q_n) + {\rm
  d}\sigma_\rho^{(1)}(\phi_n,q_n) + \int_0^{\rho}{\rm d}q_{n+1}
\frac{{\rm d}\sigma^{(0)}(\phi_{n+1},q_{n+1})}{{\rm
    d}q_{n+1}}\theta(q_n-\rho) + {\cal O}(\alpha_s^2) \right) \ .
\end{equation}
The Sudakov weight attached to the correction cross section is not
running down to the merging scale, as one would naively expect. In
this case, a double counting of logarithms of the merging scale would
have happened between the nLO cross section and the full one-loop
correction. Indeed, the role of the correction we apply here is to
precisely replace the first order virtual approximation stemming from
expanding the Sudakov form factor by its exact counterpart. Fulfilling
this condition is at the heart of the merging efforts at NLO presented
so far,
\cite{Lavesson:2008ah,Hoeche:2012yf,Gehrmann:2012yg,Frederix:2012ps}.

Let us stress the importance of maintaining inclusive cross sections
in the case of NLO corrections. The problematic contribution by which
the inclusive cross section differs from its expected value in the
case of tree level merging, as discussed in
section.~\ref{secs:inclusive}, is given by
$$
\delta\left({\rm d}\sigma_{n,\text{incl}}\right) =
\int_\rho^{q_{n}} {\rm d}q_{n+1} \left( \frac{{\rm
    d}\sigma_\rho^{(0)}(\phi_{n+1},q_{n+1})}{{\rm d}q_{n+1}}-\frac{{\rm
    d}\phi_{n+1}}{{\rm d}\phi_{n}{\rm d}q_{n+1}}
P_\rho(\phi_{n},q_{n+1}){\rm d}\sigma_\rho^{(0)}(\phi_{n},q_{n})
\right)\Delta_{n}(q_{n+1}|\cdots|q_0) \ .
$$ This cross section could well be expected not to contribute
logarithmically enhanced terms, provided the shower is a good
approximation to singly unresolved limits of tree level matrix
elements. Upon replacing the LO exclusive cross sections above the
merging scale by their NLO counterparts, the above argument does not
apply anymore owing to the case that NLO corrections to the shower
splitting kernels are not considered. We therefore expect violations
of inclusive cross sections at a level of logarithmic approximation
which is not covered by the shower anymore. Note that if corrections
are only available to the lowest order process $n=0$, then the
constraints on merged inclusive cross sections will in this case
ensure that also the inclusive NLO cross section is preserved. This is
however not the case starting from adding NLO corrections to the $n=1$
processes.

As for the tree level case, we can however provide correction terms
similar to the \textsf{LoopSim} nNLO corrections.  In
the presence of tree level matrix elements for up to $N$ additional
partons, and one-loop corrections for up to $M<N$ additional partons
we then arrive at the merged cross section,
\begin{eqnarray}
\label{eqs:nlomerged}
{\rm d}\sigma^{\text{merged}}_{N,M,\rho} &= &
\sum_{k=0}^{N-1} {\rm PS}_\rho^{-1}\left[ 
\left( {\rm d}\sigma_\rho^{(0)}(\phi_k,q_k)-\int_\rho^{q_k}{\rm d}q_{k+1}  
\frac{{\rm
  d}\sigma_\rho^{(0)}(\phi_{k+1},q_{k+1})}{{\rm d}q_{k+1}} \Delta_{k}(q_{k+1}|q_k)\right) \Delta_{k-1}(q_k|\cdots|q_0)
\right]
\\\nonumber
&+&{\rm d}\sigma_\rho^{(0)}(\phi_N,q_N)\Delta_{N-1}(q_N|\cdots |q_0)
\\ \nonumber 
&+&
\sum_{k=0}^{M-1} {\rm PS}_\rho^{-1}\left[
\left( {\rm d}\sigma_\rho^{(1,\text{incl})}(\phi_k,q_k)-\int_\rho^{q_k}{\rm d}q_{k+1}  
\frac{{\rm
  d}\sigma_\rho^{(1,\text{incl})}(\phi_{k+1},q_{k+1})}{{\rm d}q_{k+1}} \Delta_{k}(q_{k+1}|q_k)\right) \Delta_{k-1}(q_k|\cdots|q_0)
\right]
\\\nonumber
&+& {\rm PS}_\rho^{-1}\left[
{\rm d}\sigma_\rho^{(1,\text{incl})}(\phi_M,q_M)\Delta_{M-1}(q_M|\cdots |q_0)
  \right] 
\end{eqnarray}
which, upon parton shower action, will provide a merged sample with
NLO accuracy for up to $M$-jet observables, LO accuracy for up to
$N$-jet observables, including resummation at whatever accuracy is
provided by the parton shower. Here, we have denoted the NLO
corrections to inclusive cross sections as
\begin{equation}
{\rm d}\sigma_\rho^{(1,\text{incl})}(\phi_n,q_n) =
{\rm d}\sigma_\rho^{(1)}(\phi_n,q_n)
  + \int_0^{q_n}{\rm d}q_{n+1}\frac{{\rm
      d}\sigma^{(0)}(\phi_{n+1},q_{n+1})}{{\rm
      d}q_{n+1}}\theta(q_n-\rho) \ .
\end{equation}
Note that the action of ${\rm PS}_\rho^{-1}$
integrates to one, which clarifies once more that we are able to
preserve inclusive cross sections.
Given the extensive discussions on reclustering, dynamic scale choices
and the generation of the Sudakov weights in the context of merging
approaches so far\cite{Lonnblad:2001iq,Krauss:2002up,Hoche:2006ph,
  Lavesson:2007uu,Hoeche:2009rj,Hamilton:2009ne,Lonnblad:2011xx}, we
will not include a detailed algorithmic definition of the merging
procedure here. Technical aspects will be subject to ongoing and
future work concerned with the implementation of the procedure
outlined here.

\section{Corrections below the merging scale}
\label{secs:below}

Having derived the merged cross section in the presence of both tree
level and one loop matrix elements, note that we actually have solved
a NLO matching condition for each exclusive $n$-parton contribution above
the merging scale, cf. eq.~\ref{eqs:mergematch}. More precisely,
\begin{itemize}
\item by preserving inclusive cross sections in the presence of only
  tree level matrix elements, we have matched each exclusive
  $n$-parton contribution to a \textsf{LoopSim} approximated nLO, and
\item by including exact NLO corrections, we have fulfilled this
  matching condition at NLO, while finally
\item by preserving inclusive cross sections in the latter case, we
  start to generate approximate NNLO pieces, which could well be the
  basis for NNLO matching.
\end{itemize}
These considerations apply to contributions with emissions above the
merging scale. Below the merging scale, we are still left with the
shower approximation, and we will finally give a simple prescription
of how nLO (in the case of tree level merging) and NLO (in the case of
one loop merging) accuracy can be achieved also below the merging
scale. To be precise, we consider the cases of $n$ partons above
the merging scale and the first order contribution to zero or one
emission below the merging scale.
Demanding the required nLO or NLO accuracy, we obtain generalizations
of the familiar NLO matching corrections restricted to the phase space
for one emission below the merging scale,
\begin{align}
{\rm d}\sigma_{N,\mu|\rho}^{\text{match}} &= \sum_{k=0}^{N-1}{\rm PS}^{-1}_\rho\left[ 
\left({\rm d}\sigma_{\mu|\rho}^{(0),\text{match}}(\phi_k,q_k) -
\int_\mu^\rho {\rm d}q_{k+1} \frac{{\rm d}\sigma_{\mu|\rho}^{(0),\text{match}}(\phi_{k+1},q_{k+1})}{{\rm d}q_{k+1}}
\Delta_k(q_{k+1}|q_k)\right)\Delta_{k-1}(q_k|\cdots|q_0)
\right]\\\nonumber
&+ {\rm d}\sigma_{\mu|\rho}^{(0),\text{match}}(\phi_N,q_N) \Delta_{N-1}(q_N|\cdots|q_0)
\end{align}
with
\begin{equation}
{\rm d}\sigma_{\mu|\rho}^{(0),\text{match}}(\phi_n,q_n) =
{\rm d}\sigma_{\mu}^{(0)}(\phi_n,q_n)\theta(q_{n-1}-\rho)\theta(\rho-q_n)
-{\rm d}\sigma_\rho^{(0)}(\phi_{n-1},q_{n-1})\frac{{\rm d}\phi_n}{{\rm d}\phi_{n-1}}
P_{\mu|\rho}(\phi_{n-1},q_n) \ .
\end{equation}
Note that these corrections do not change inclusive cross sections as
is the case for plain NLO matching. Within these contributions, the
parton shower cutoff can be sent to zero provided the singly
unresolved limits are reproduced properly; a finite shower cutoff will
act similar to a phase space slicing parameter in terms of which the
NLO cross section is reproduced.

\section{Conclusions and outlook}
\label{secs:conclusions}

We have presented an extension to multileg matrix element and parton
shower merging, which preserves inclusive cross sections at the level
of the available accuracy, particularly at tree and one-loop
level. This constraint is of utmost importance particularly for the
latter case, as NLO corrections to shower splitting kernels are so far
out of reach. This lack of shower accuracy manifests itself in terms
of potentially large logarithmic contributions which are of the same
order of magnitude as the NLO corrections tackled in recent approaches
of combining fixed order corrections and parton shower resummation,
thus spoiling NLO accuracy for lower jet multiplicities.

The formalism used to derive the modified algorithm is general enough
to study the inclusion of even higher order corrections by
successively replacing virtual contributions as approximated by the
shower through their exact counterpart. The ingredients at for tree
level and one loop merging, respectively, are given by fully
differential LO calculations and {\it inclusive} NLO corrections
differential in the respective Born variables. An implementation of
the algorithm presented is subject to ongoing and future work.

\section*{Note added}

During completion of this work a very similar approach to the problem
by L\"onnblad and Prestel came to the attention of the author, first
results of which have been presented for the tree level case in
\cite{Lonnblad:2012ng} and for the one-loop case in
\cite{Lonnblad:2012ix}.

\section*{Acknowledgments}
This work was supported by the Helmholtz Alliance ``Physics at the
Terascale''. The author wants to thank Johannes Bellm, Leif
L\"onnblad, Stefan Prestel and Frank Tackmann for fruitful discussion
on the subject. The author acknowledges the kind hospitality of the
Galileo Galilei Institute for Theoretical Physics where first studies
connected to the present work have been carried out.

\bibliography{multijets}

\end{document}